# ac strain based thermodynamic criterion for vortex lattice in type-II superconductors

Peipei Lu[1,2,§], Mengju Yuan[1,§], Jing Zhang[1], Qiang Gao[3], Shuang Liu[4], Yugang Zhang[1], Shipeng Shen[3], Long Zhang[1], Jun Lu[3], Xiaoyuan Zhou[4], Mingquan He[1], Aifeng, Wang[1], Yang Li[3,5], Wenshan Hong[3], Shiliang Li[3], Huiqian Luo[3], Xingjiang Zhou[3], Xianhui Chen[6], Young Sun[3,4,*] and Yisheng Chai[1,4,†]

[1]Low Temperature Physics Laboratory, College of Physics, Chongqing University, Chongqing 401331, China.

[2]College of Physics and Hebei Advanced Thin Films Laboratory, Hebei Normal University, Shijiazhuang 050024, Hebei, China.

[3]Beijing National Laboratory for Condensed Matter Physics, Institute of Physics, Chinese Academy of Sciences, Beijing 100190, China.

[4]Center of Quantum Materials and Devices, Chongqing University, Chongqing 401331, China.

[5]University of Chinese Academy of Sciences, Beijing 100190, China.

[6]CAS Key Laboratory of Strongly-coupled Quantum Matter Physics, Department of Physics, University of Science and Technology of China, Hefei, Anhui, China.

[*]Corresponding authors. youngsun@iphy.ac.cn

[†]Corresponding authors. yschai@cqu.edu.cn

[§]These authors contributed equally to this work.

We report a dynamic magnetostrictive effect in type-II superconductors (e.g., Nb, $YBa_2Cu_3O_{7-x}$, $Bi_2Sr_2CaCu_2O_{8+\delta}$ and $Ba_{0.6}K_{0.4}Fe_2As_2$) and any type-II superconductor is expected to present a similar response to this new technique. Measured via a composite magnetoelectric technique, an ac field excites an in-phase, non-dissipative strain response scaling linearly with vortex density in the vortex lattice phase. In the vortex liquid phase, the signal acquires an out-of-phase component before vanishing in the normal state. We propose the ac strain susceptibility a thermodynamic criterion for the vortex-lattice, capturing vortex collective modes inaccessible from static measurements.

# I. INTRODUCTION

Superconductors are defined by two hallmark properties: zero electrical resistivity and perfect diamagnetism, known as the Meissner effect [1,2]. These properties persist until superconductivity is quenched by a magnetic field exceeding a critical threshold $H_c$ in type-I superconductors. By comparison, in type-II superconductors, the negative interfacial energy between superconducting and normal regions allows magnetic flux to penetrate the material above a lower critical field $H_{c1}$. This flux forms discrete vortices, each carrying a single flux quantum $\phi_0 = hc/2e$ (where $h$ is the Planck constant and $e$ is the electron charge). The quantum nature of these vortices facilitates a range of emergent phenomena, from their potential use as superconducting qubits to hosting Majorana zero modes in topological quantum computing [3,4]. These vortices typically arrange into regular two-dimensional (2D) lattices—often triangular or square in symmetry [5]. As temperature ($T$) or magnetic field increases, the rigid vortex lattice can melt into a dissipative vortex-liquid state above the irreversibility line ($T_{irr}$ or $H_{irr}$), ultimately transitioning to the normal state at the upper critical field ($H_{c2}$).The static and dynamic properties of these phases play a central role in shaping the magnetic and transport responses of the material. For instance, hysteresis in magnetization loops $M(H)$ and the divergence between zero-field-cooled (ZFC) and field-cooled (FC) magnetization curves are classical manifestations of vortex pinning [6]. Similarly, the critical current density $J_c$, is determined by the ability of the vortex lattice to remain static under an applied current [7], while, the Nernst effect provides a hallmark signature of vortex mobility in the vortex-liquid regime [8-10].

Complementary to these macroscopic probes, the thermodynamic strain channel provides a complementary probe of vortex phases via magnetostriction ($\lambda = \Delta L/L$, where $L$ is the sample dimension). In type-II superconductors, static magnetostriction $\lambda_{dc}$ arises from magnetoelastic coupling between the vortex system and the parent crystal lattice. It can be characterized using conventional strain gauges or capacitive dilatometers, often revealing butterfly-shaped hysteresis loops or discontinuous jumps due to thermomagnetic flux avalanches [11-15]. However, while such measurements only capture the quasi-static state, they do not easily isolate collective excitation mode [16,17] or dissipation behavior within the vortex phases.

In contrast, the magnetic strain susceptibility, ($d\lambda/dH$), opens a window onto the dynamical aspects of the vortex system. To formalize this, we consider the Gibbs free energy G($H$,$\sigma$,$T$), where σ is stress. The field derivative of strain, $(\partial\lambda/\partial H)_{\sigma,T}$, is a mixed second derivative of the free energy (a Maxwell relation) and thus a bona fide strain susceptibility linked to magnetization via $(\partial\lambda/\partial H)_{\sigma,T}=-(\partial M/\partial\sigma)_{H,T}$. Historically, $d\lambda/dH$ has been obtained by differentiating $\lambda(H)$ data from dc measurements, a method cannot probe dynamic and dissipative contributions [18]. Under a small ac drive, however, this susceptibility is complex:

$$(d\lambda/dH)_{ac} = d\lambda'/dH + i d\lambda''/dH \quad (1)$$

where the real part $d\lambda'/dH$ represents an elastic response, while the imaginary part $d\lambda''/dH$ quantifies dissipation. Accessing this complex susceptibility therefore offers a practical strategy to probe the dynamic/dissipative properties of vortex phases.

It is only recently that this complex quantity has been measured with ultrahigh precision using a composite magnetoelectric (ME) technique. This method has been successfully applied to study the dynamics of quantized skyrmion lattices, other magnetically ordered systems and even quantum oscillation in topological semimetals [19-25]. The ME technique relies on a mechanically bonded heterostructure where length oscillations driven by a small ac field ($H_{ac} \approx 0.5 - 1$ Oe; $f \leq 10$ kHz) are converted into a measurable ac voltage ($V_{ac}$) by an attached piezoelectric layer (Fig. 1a). A lock-in amplifier reads the voltage at high speeds (1-5 point/s), providing a significant advantage over conventional ac susceptibility. As the generated voltage scales with the complex transverse magnetostrictive coefficient ($V_{ac} \sim (d\lambda/dH)_{ac}$), we treat the measured signal as a direct proxy for the thermodynamic strain susceptibility.

In this letter, we report the observation of a dynamic magnetostrictive effect in archetypal type-II superconductors, including Nb, $YBa_2Cu_3O_{7-x}$ polycrystals and $Bi_2Sr_2CaCu_2O_{8+\delta}$, $Ba_{0.6}K_{0.4}Fe_2As_2$ single crystals. We reveal a robust linear correlation between the amplitude of deformation and the vortex density in the lattice phase across all systems. Notably, we demonstrate that this dynamic effect is distinct from the static magnetostriction, evidenced by a stark contrast between their respective coefficients during flux avalanches and at the irreversibility line. These features establish the ac strain susceptibility as a robust thermodynamic criterion for identifying the vortex lattice state.

## II. METHODS

Commercial polycrystalline Nb and $YBa_2Cu_3O_{7-x}$ (YBCO) polycrystals samples were obtained from Alfa Aesar and the Central Iron & Steel Research Institute (China), respectively. Optimally doped and $Bi_2Sr_2CaCu_2O_{8+\delta}$ (BSCCO) and $Ba_{0.6}K_{0.4}Fe_2As_2$ (BKFA) single crystals were synthesized using the traveling solvent floating-zone method and the self-flux method, respectively [26,27]. Resistance was measured using a standard four-probe configuration. Magnetization data were collected using the Vibrating Sample Magnetometer (VSM) module of the PPMS system. Ac magnetic susceptibility $\chi = \chi'$-$i\chi''$ was measured for Nb, YBCO, BSCCO, and BKFA samples using the ACMS option of the PPMS, with variable excitation frequencies. Dc magnetostriction is characterized with a commercial CuBi dilatometer (MultiField Tech.) with high precision capacitance bridge AH2700 (Andeen-Hagerling). Composite magnetoelectric (ME) structures were prepared by bonding each type-II superconductor to a piezoelectric $0.7Pb(Mg_{1/3}Nb_{2/3})O_3$-$0.3PbTiO_3$ (PMN-PT) [001]-cut single crystal (thickness = 0.2 mm) using silver epoxy (Epo-Tek H20E, Epoxy Technology Inc.). Prior to electrical measurements, the PMN-PT substrates were poled along the thickness direction using an electric field of 5.5 kV/cm for 1 hour at room temperature. $V_{ac}$ was measured using a lock-in amplifier (NF Corporation LI5645) integrated with a commercial sample stick (MultiField Tech.), as illustrated in Fig. 1a. All the $d\lambda'/dH$ values were corrected by subtracting a small background signal from the normal state, which was either constant or linearly dependent on $H_{dc}$.

## III. RESULTS AND DISCUSSION

### A. Dynamics magnetostrictive effect in polycrystalline Nb

We begin with the investigation of the classical low-$T_c$ superconductor Nb. The superconducting transition was characterized using conventional resistance ($R$) and dc magnetization measurements. As shown in Fig. 1b, the transition temperature ($T_c$) is approximately 9.0 K at zero magnetic field, consistent with literature values [28,29]. $T_c$ is progressively suppressed by the magnetic field, reaching ~3 K at 1 T. This is corroborated by $M(T)$ curves (see Supplemental Material (SM) for details) [30]. Under a field of 0.02 T, a pronounced separation between ZFC and FC magnetization curves indicates strong diamagnetism (superconducting volume fraction ~ 68.8%, Fig. 1c). The convergence of ZFC and FC curves at $T_{irr}$ = 8.8 K marks the transition from the frozen vortex-lattice to the mobile vortex-liquid phase.

To investigate the dynamic magnetostrictive properties, a Nb/PMN-PT composite structure was fabricated. Figure 1d presents the temperature dependence of $(d\lambda/dH)_{ac}$ measured during FC processes. In the absence of a dc magnetic field, the signal remains negligible zero both below and above $T_c$, indicating no dynamic magnetostriction without vortex formation. However, when a finite magnetic field (up to 0.7 T) is applied, the real part $d\lambda'/dH$ exhibits a sharp drop at $T_c$, forming a pronounced negative plateau at lower temperatures. Simultaneously, the imaginary part $d\lambda''/dH$ displays a single dip centered around the transition region (see SM [30]). Both features are gradually suppressed with increasing field, vanishing completely near the upper critical field $H_{c2}$~0.8 T [32].

The emergence of a non-zero $d\lambda''/dH$ near $T_c$ is attributed to dissipation in the vortex-liquid phase (see SM [30]). To validate this, we measured the ac magnetic susceptibility $\chi = \chi' - i\chi''$ under $\mu_0 H_{dc}$ = 0.5 T (see SM [30]). The temperature range of the $d\lambda''/dH$ dip coincides precisely with the peak in the magnetic loss signal $\chi''$. This supports the assignment of the imaginary strain response to dissipative vortex dynamics due to depinning. Accordingly, we therefore use the upper and lower bounds of the $d\lambda''/dH$ dip as operational estimates of $T_c$ and $T_{irr}$, respectively, and compare them with resistance and magnetization criteria (see SM [30]). This criterion is further verified by comparing ZFC and FC $(d\lambda/dH)_{ac}$ curves (Fig. 1e). At low temperatures, clear differences emerge due to flux trapping, but the curves converge precisely at $T_{irr}$, the onset of the dissipative dip (see SM [30]).

Detailed analysis of the d$\lambda'$/d$H$ plateau at 4 K reveals that its magnitude increases nearly linearly with magnetic field strength up to 0.7 T (Fig. 1f). The finite d$\lambda'$/d$H$ in the vortex-lattice phase cannot arise solely from the applied background field $H_{dc}$, as the ZFC and FC results differ markedly. A more plausible interpretation is that the magnitude of d$\lambda'$/d$H$ directly reflects the vortex density. This hypothesis is supported by the ZFC data, which shows an increase in signal magnitude with rising temperature as thermal fluctuations facilitate the entry of vortices into the lattice.

To distinguish the dynamic response from static deformation, we performed field-dependent measurements on Nb at 2.5 K (Fig. 2). The magnetization $M(H)$ exhibits strong diamagnetism and deviates from linearity at $H_{c1}\sim$ 0.13 T. Above $H_{c1}$, the loops display pronounced hysteresis and discrete magnetization jumps—flux avalanches—large number of vortices rush in or out of the sample in a very short time (see inset of Fig. 2a). These avalanches persist up to 8.5 K (see SM [30]). Above $H_{irr}$, the hysteresis vanishes. In the ac magnetostrictive measurements (Fig. 2b), the real part d$\lambda'$/d$H$ does not exhibit a clear anomaly at $H_{c1}$. Instead, it tracks the vortex density: during vortex avalanches, d$\lambda'$/d$H$ exhibits step-like increases and decreases, while the imaginary part d$\lambda''$/d$H$ remains zero. The amplitude of d$\lambda'$/d$H$ peaks at $H_{irr}$, where the vortex density in lattice phase is maximal. At higher fields ($H > H_{irr}$), where $M(H)$ is nearly zero, d$\lambda'$/d$H$ rapidly decreases, vanishing near $H_{c2}$. In the intermediate range between $H_{irr}$ and $H_{c2}$, the d$\lambda''$/d$H$ signal exhibits a single dip, consistent with the dissipative vortex-liquid phase observed in temperature sweeps.

The dc transverse magnetostriction $\lambda_{dc}$ (Fig. 2c) displays a butterfly-shaped loop characteristic of type-II superconductors [11,13-15], with abrupt jumps matching the vortex avalanches. However, the derivative of this static strain, d$\lambda_{dc}$/d$H$ (Fig. 2d), reveals a fundamental difference. The d$\lambda_{dc}$/d$H$ are dominated by sharp, singular peaks during avalanches and vanish entirely at $H_{irr}$. In contrast, the dynamic coefficient d$\lambda'$/d$H$ shows monotonic steps during avalanches and reaches its maximum magnitude at $H_{irr}$. This pronounced discrepancy proves that the two measurements probe distinct physical phenomena. The dc coefficient reflects macroscopic, quasi-static flux redistribution (which stops at $H_{irr}$), whereas the ac coefficient probes the intrinsic, collective elasticity of the pinned vortex lattice itself, which is maximally ensembles just before melting.

B. Dynamics magnetostrictive effect in other high-Tc superconductors

To test the generality of this phenomenon, we extended our study to three additional archetypal superconductors: YBCO (Cu-based polycrystal), BSCCO (Cu-based single crystal), and BKFA (Fe-based single crystal). Resistance measurements yield transition temperatures of $T_c$ = 86.4, 87.2, and 38.5 K, respectively (Fig. 3a-c). The $R(T)$ curve for YBCO displays two steps, hinting at a vortex-slush phase [33]. Magnetization measurements confirm the type-II nature and the boundaries of the lattice phase (Fig. 3d-f). Composite ME structures were fabricated for each material. The temperature dependence of $d\lambda'/dH$ (Fig. 3g-i) and $d\lambda''/dH$ (see SM [30]) was measured under selected dc fields. In all cases, $d\lambda''/dH$ reveals one or more dips, signaling the presence of dissipative phases. These dips coincide with peaks in the ac magnetic susceptibility $\chi''$, consistent with the observations in Nb (see SM [30]). Below $T_{irr}$, where $d\lambda''/dH$ vanishes, the real part $d\lambda'/dH$ exhibits a pronounced negative plateau (YBCO, BKFA) or a weakly temperature-dependent behavior (BSCCO). Specific material features are clearly resolved by the technique. In YBCO, an additional low-temperature peak in $\chi''$ appears consistent with vortex shaking or depinning induced by the strong ac field. In BSCCO, the weak temperature dependence of $d\lambda'/dH$ reflects the weak pinning force, allowing thermal fluctuations to increase vortex penetration at higher temperatures. In BKFA, multiple dips in $d\lambda''/dH$ near $T_{irr}$ reflect complex vortex dynamics, likely the coexistence of elastic and plastic vortex segments [34]. Despite these specific differences, the general phenomenology—a dissipative dip at the liquid phase and an elastic response in the lattice phase—is universal.

Furthermore, field-dependent measurements (Fig. 3j-l) reconfirm the linear relationship between $d\lambda'/dH$ and vortex density. Both BSCCO and BKFA exhibit an almost linear increase of the $d\lambda'/dH$ amplitude with applied field, accompanied by minimal hysteresis. YBCO displays pronounced hysteresis consistent with stronger pinning in the polycrystalline form. Crucially, these behaviors differ fundamentally from the butterfly-shaped $d\lambda_{dc}/dH$ expected from static magnetostriction [11,13-15], reinforcing the distinct origin of the dynamic signal. The systematic observation of this effect across four distinct superconductors suggests a common vortex-related origin. We model this as a collective excitation of the vortex lattice.

We expect that a small variation in the external magnetic field $\delta H$ induces two superimposed responses in crystal lattice (Fig. 4). First, a change in vortex density leads to a quasi-static deformation $\delta\lambda_{dc}$ via the interaction between vortices and pinning centers [2]. Second, rather than triggering depinning, each vortex line oscillates about its equilibrium position. Owing to boundary constraints (sample length $L$) and the dense packing of the lattice, a standing-wave-like collective mode with frequency $\omega_0$ is excited. Prior microwave experiments place an upper limit on this mode of $\omega_0/2\pi \sim 10$ GHz in the absence of depinning [35]. Consequently, the measured ac magnetostrictive coefficient can be decomposed into two terms:

$$(d\lambda/dH)_{ac}=\delta\lambda_{dc}/H_{ac} + \delta\lambda_{ac}/H_{ac}. \quad (2)$$

The first term reflects the static density change. The second, dynamic term can be approximated as:

$$\delta\lambda_{ac}/H_{ac}\approx \frac{ng}{\omega_0^2} \quad (3)$$

where $n$ is the total number of vortices and $g$ is the effective acceleration factor per vortex line (see SM for derivation [30]). For a typical sample area of 1 $mm^2$ under 1 T, the vortex number $n$ reaches $10^9$. With $\omega_0$ in the GHz range, the term $ng/\omega_0^2$ would yield a measurable elastic response. This model successfully accounts for our key observations: (1) the large magnitude of the dynamic signal compared to the static derivative; (2) the linear scaling with vortex density ($n \sim \mu_0 H_{dc}$); and (3) the purely elastic nature (zero phase lag) within the lattice phase where the drive frequency (<10 kHz) is far below the resonant frequency $\omega_0$. In the vortex-liquid phase, vortex depinning, introducing dissipation and a finite imaginary component $d\lambda''/dH$. In the normal state ($n$=0), the signal vanishes. This theoretical framework explains why the dynamic strain susceptibility provides information of the lattice complementary to static measurements.

From a practical standpoint, the measurement is rapid, robust, and requires only simple instrumentation, providing an accessible route to map vortex phase diagrams and quantify magnetoelastic coupling. To demonstrate its broader capability, we also applied this approach to the type-II superconductor $EuFe(As_{1-x}P_x)_2$, where superconductivity coexists with ferromagnetism. In this technique, two phases provide signals in opposite signs. For samples with $T_c<T_{FM}$ ($T_{FM}$ is ferromagnetic transition temperature), the transition to a ferromagnetic multi-domain state suppresses $H_{c2}$ and $H_{irr}$. Conversely, for samples with $T_c>T_{FM}$, strong short-range spin correlations and phase boundaries within a multiphase coexistence regime near a triple point leads to pronounced $H_{c2}$ suppression [36].

## IV. CONCLUSIONS

In conclusion, we have established the dynamic magnetostrictive effect as a strain-based thermodynamic criterion for the vortex lattice state, demonstrating a linear scaling with vortex density in type-II superconductors. Unlike static magnetostriction, the ac strain susceptibility isolates the collective elasticity of the pinned lattice from the dissipative liquid. This technique provides a rapid, sensitive probe of vortex matter, offering a powerful complement to transport and magnetization measurements for mapping phase diagrams of type-II superconductors.

## ACKNOWLEDGEMENTS

This work was supported by the National Natural Science Foundation of China (Grant Nos. 12227806, 12374081, 11674347, 11974065, 51725104, 11774399, 11474330, 52101221, U21A201910), Fundamental Research Funds for the Central Universities (Project No. 2024IAIS-ZX002), the National Key Research and Development Program of China (Grants No. 2023YFA1406100), the Central Guidance on Local Science and Technology Development Fund of Hebei Province (Grant No. 246Z7611G), Science Research Project of Hebei Education Department (Grant No. BJ2025091) and the Open Research Fund of the Pulsed High Magnetic Field Facility (Grant No.WHMFC2024007), Huazhong University of Science and Technology. Y. S.

Chai would like to thank the support from Beijing National Laboratory for Condensed Matter Physics. We would like to thank Miss G. W. Wang and Y. Liu at Analytical and Testing Center of Chongqing University for their assistance. We would like to thank Hengyu Guo for his assistance in drawing the schematics.

DATA AVAILABILITY

The data that support the findings of this article are not publicly available upon publication because it is not technically feasible and/or the cost of preparing, depositing, and hosting the data would be prohibitive within the terms of this research project. The data are available from the authors upon reasonable request.

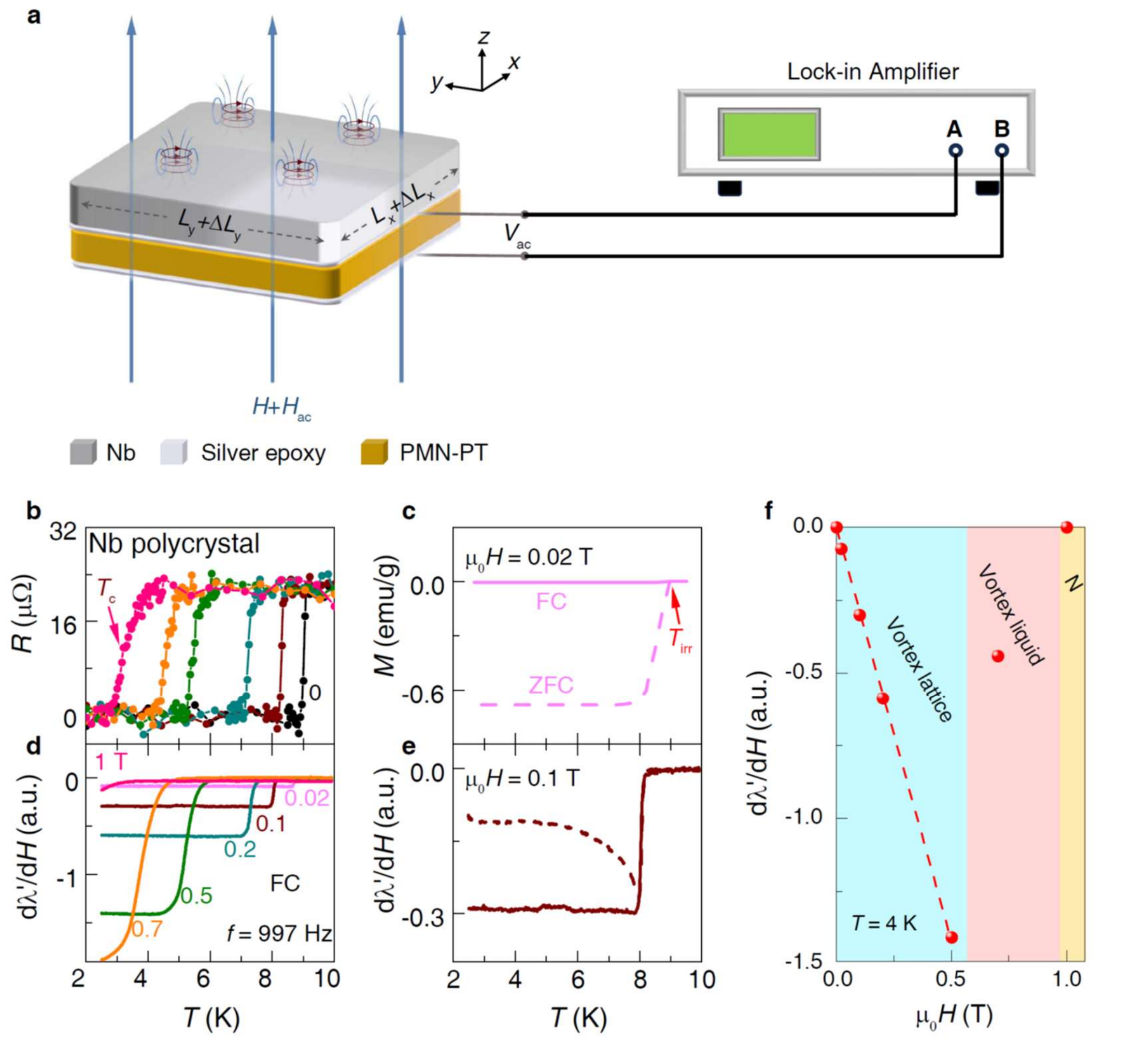

**Fig. 1. a**, Schematic diagram of the Nb/PMN-PT composite configuration used for dynamic magnetostrictive measurements. $T$ dependent **b**, $R$ under various $H$ **c**, magnetization in FC and ZFC processes under 0.02 T, **d**, dλ'/d$H$ in FC process under selected $H$, and **e**, dλ'/d$H$ in ZFC and FC processes under 0.1 T for Nb polycrystal. **f**, $H$ dependenct dλ'/d$H$ at 4 K, extracted from **d**.

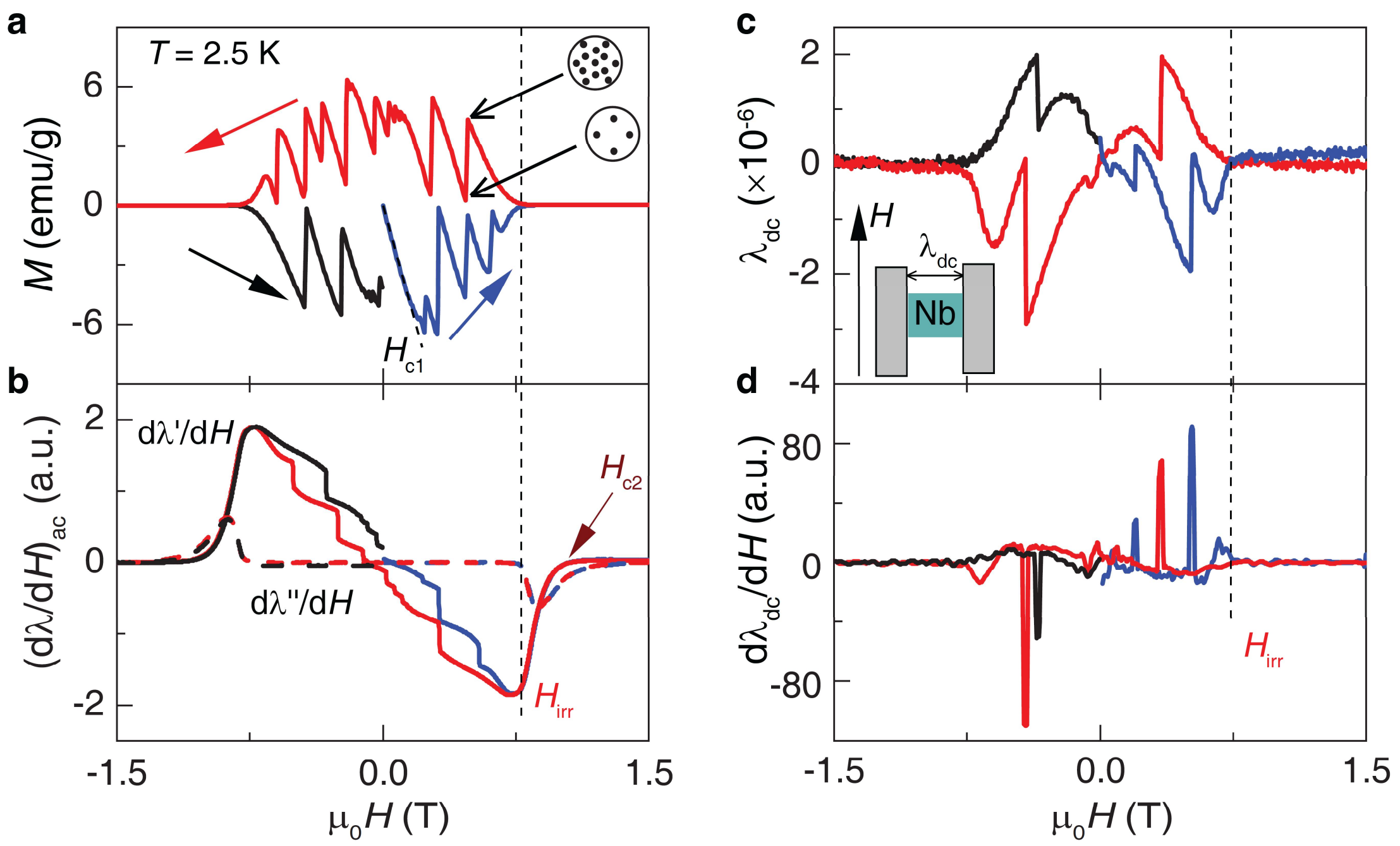

**Fig. 2.** *H* dependent **a**, *M*, **b**, dλ'/d*H* and dλ''/d*H*, **c**, transverse dc magnetostiction $\lambda_{dc}$, and **d**, $d\lambda_{dc}/dH$ at 2.5 K for Nb polycrystal. The inset of **a** shows a schematic of a vortex avalanche event. The inset of **c** shows the measurement configuration for $\lambda_{dc}$. All measurements were performed after ZFC to 2.5 K.

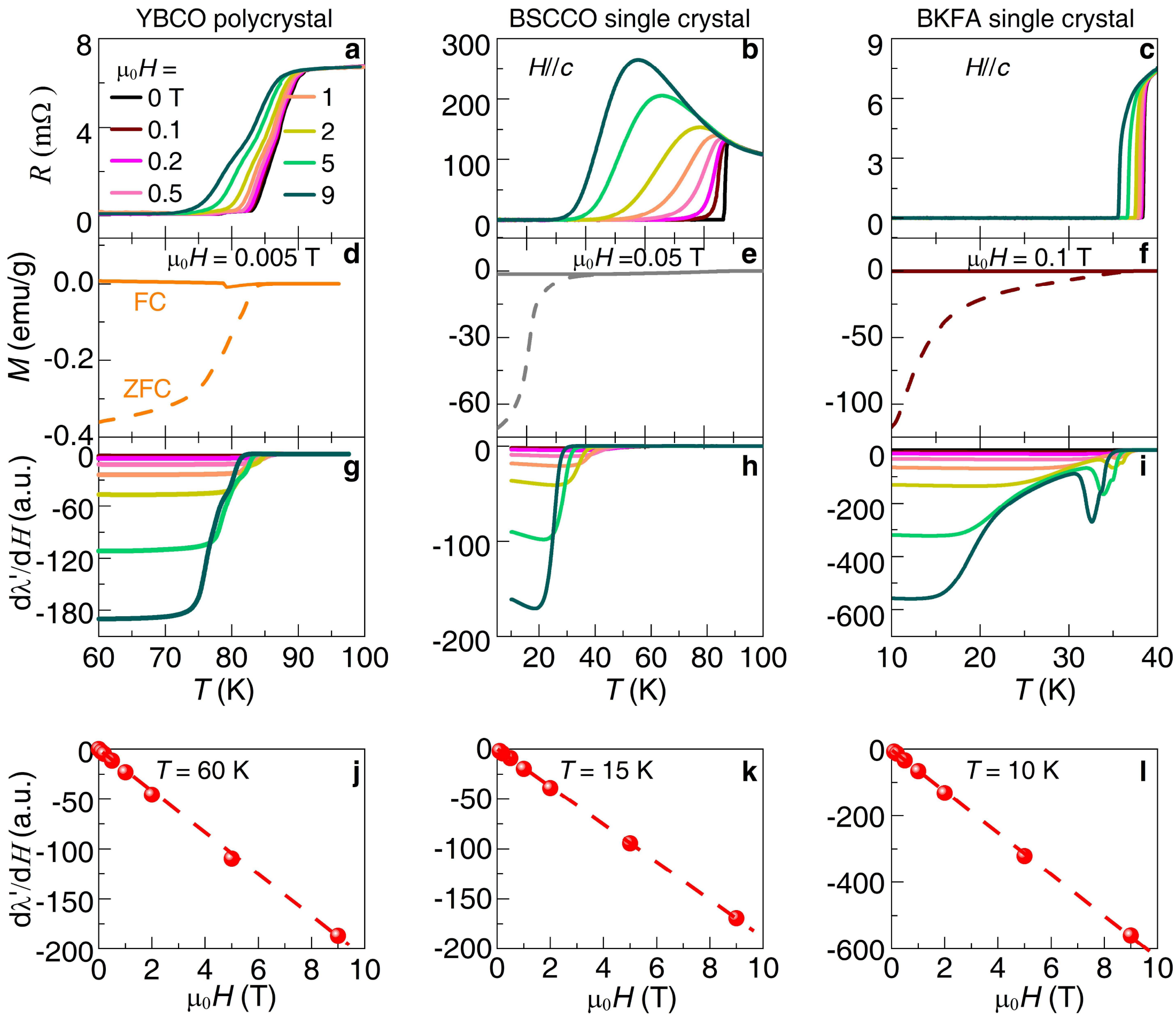

**Fig. 3.** **a-c**, $R(T)$ in FC under selected $H$; **d-f**, $M(T)$ measured during FC and ZFC under magnetic fields of **d**, 0.005 T (YBCO); **e**, 0.05 T (BSCCO); **f**, 0.1 T (BKFA); **g-i**, $d\lambda'/dH(T)$ in the FC process under selected $H$; **j-l**, $H$ dependent $d\lambda'/dH$ at selected $T$ from data in **g** (YBCO), **h** (BSCCO), and **i** (BKFA).

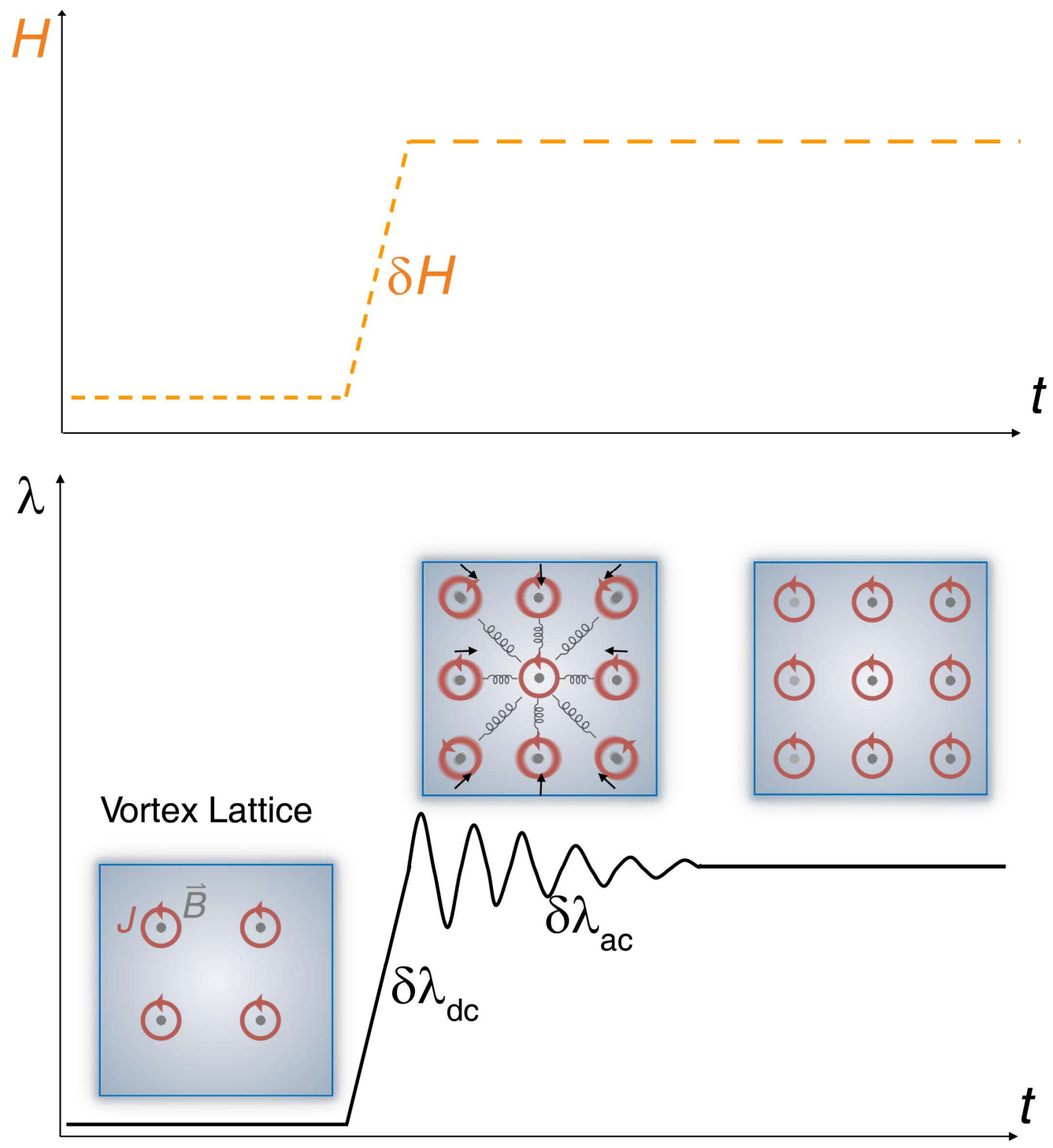

**Fig. 4.** A small variation of the external magnetic field $\delta H$ excites collective oscillations of the vortex lattice without depinning, generating a dynamic strain response $\delta\lambda_{ac}$ that is superimposed on the static deformation $\delta\lambda_{dc}$ arising from the change in vortex density.

# Supplementary Materials:

## ac strain based thermodynamic criterion for vortex lattice in type-II superconductors

In this Supplementary Materials, we provide (1) additional figures of Nb, YBCO polycrystal and BSCCO, BKFA single crystal; (2) specific explanations for $v$, and subsequently $\omega_0$; (3) Specific formula derivation for the dynamic term.

### Additional figures of Nb, YBCO polycrystal and BSCCO, BKFA single crystal

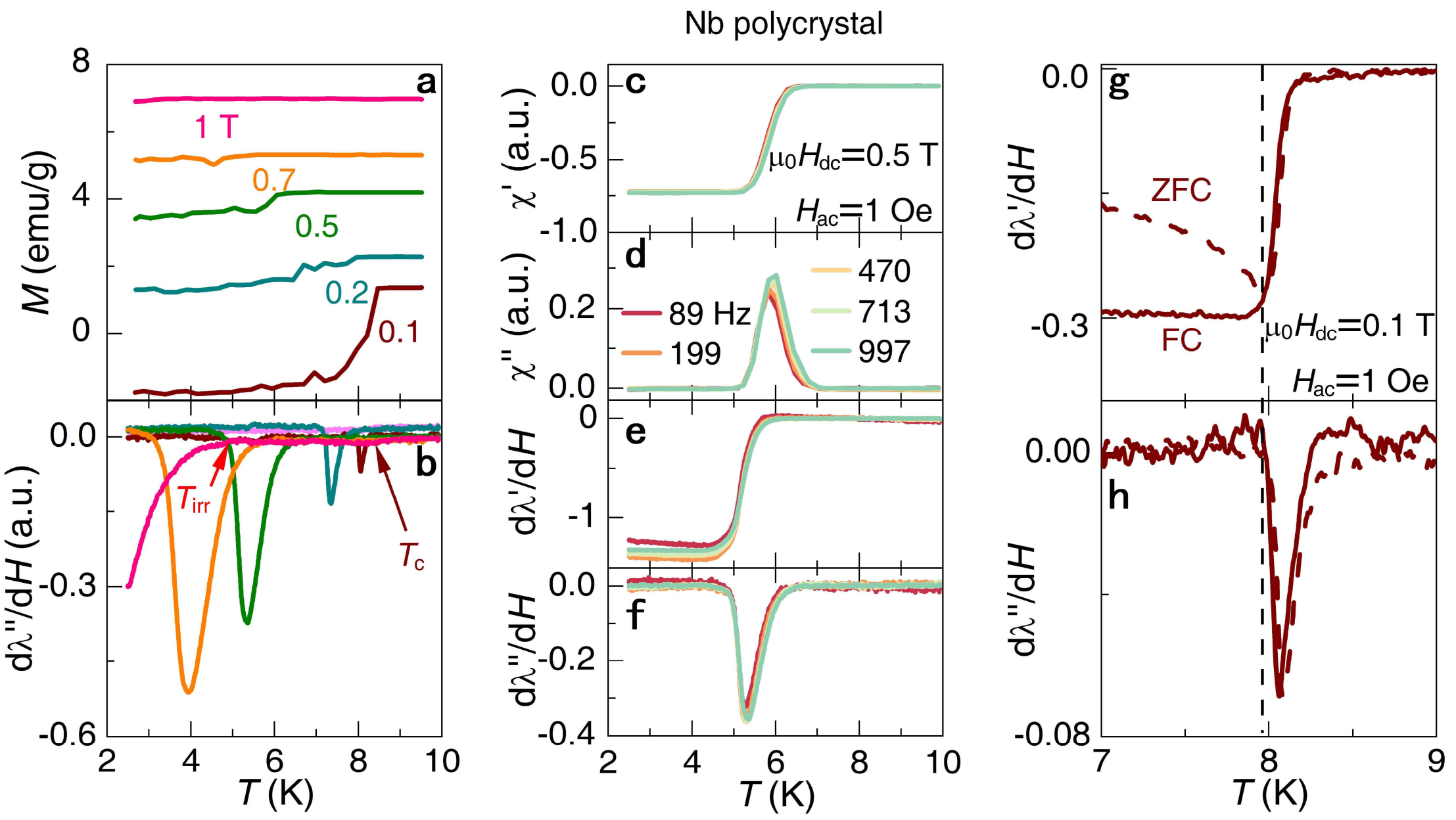

**Supplementary Figure S1. Basic characterization of Nb polycrystal.** Temperature dependence of **a**, magnetization and **b**, dλ''/d$H$ in FC process under selected magnetic fields, and **c**, real and **d**, imaginary part of the ac susceptibility χ' and χ'', respectively, and **e**, dλ'/d$H$ and **f**, dλ''/d$H$ for FC process under dc and ac magnetic fields of 0.5 T and 1 Oe, respectively, with selected frequencies. Temperature dependence of **g**, dλ'/d$H$ **h**, dλ''/d$H$ in ZFC and FC processes under 0.1 T.

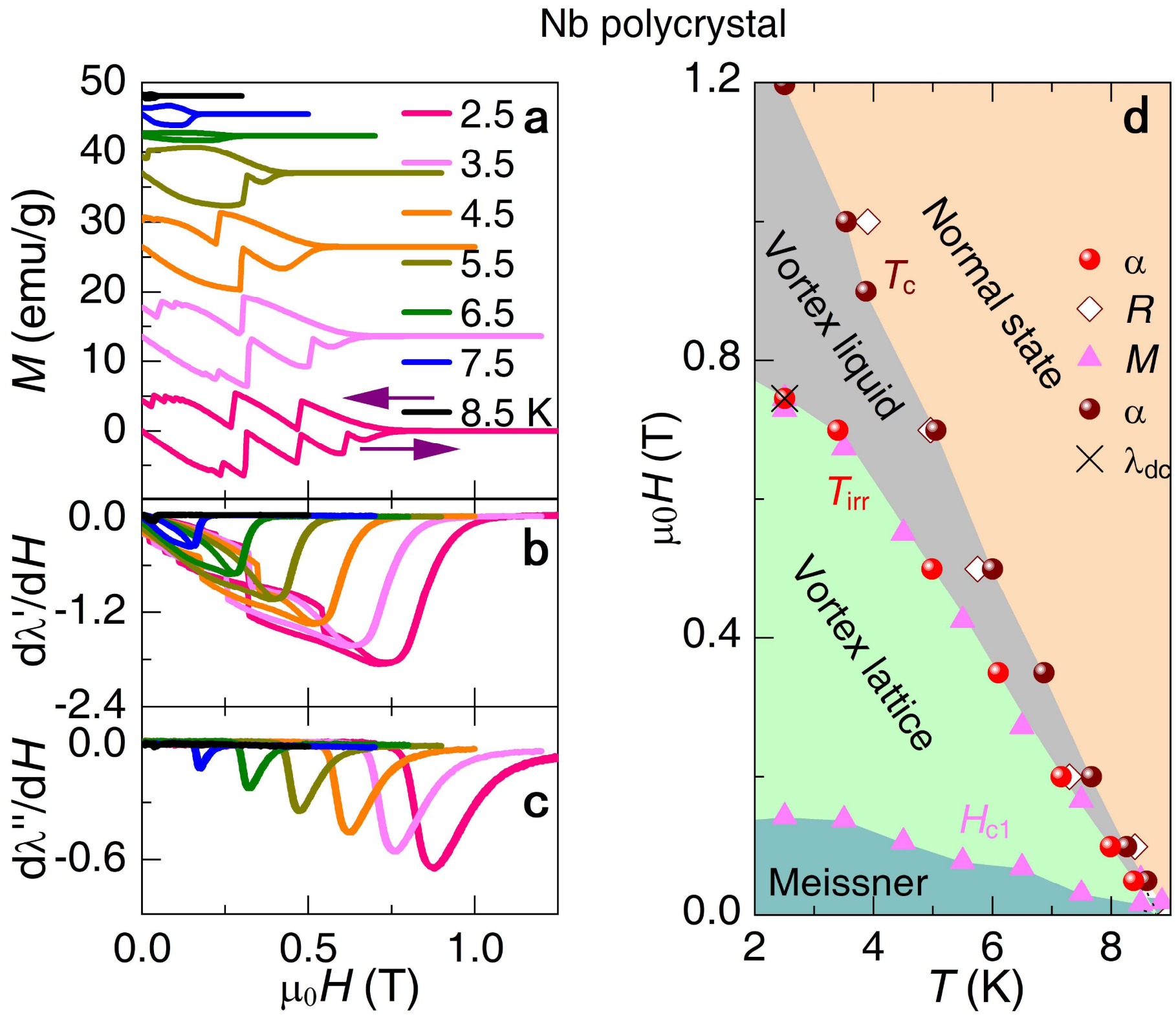

**Supplementary Figure S2. Basic characterization and *H*-*T* phase diagram of Nb polycrystal.** Magnetic field dependence of **a**, *M*, **b**, dλ'/d*H* and **c**, dλ''/d*H* at selected temperatures. All measurements were performed after ZFC to specific temperature. **d**, The *H*-*T* phase diagram based on the data in panels (S1**a** and S1**b**), (**a** to **c**), Fig. 1 and Fig. 2.

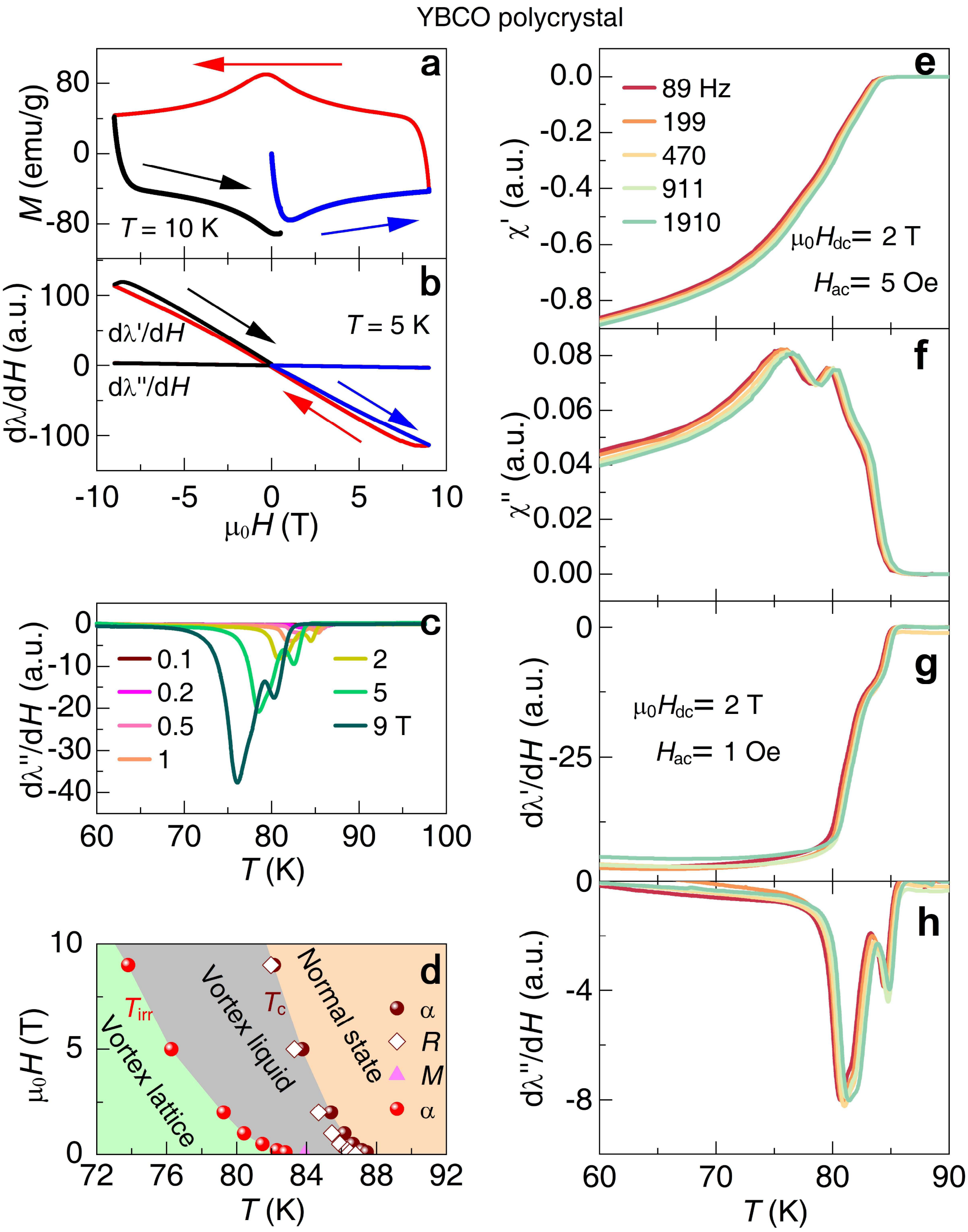

**Supplementary Figure S3. Basic characterization and *H-T* phase diagram of YBCO polycrystal.** Magnetic field dependent **a**, $M$ at 10 K and **b**, dλ'/d$H$ and dλ''/d$H$ at 5 K. Measurements were performed after ZFC process. **c**, Temperature dependenct dλ''/d$H$ in FC process under selected magnetic fields. **d**, The $H$-$T$ phase diagram based on the data in panel **c** and Fig. 3**a**, **d** and **g**. Temperature dependent **e**, χ' and **f**, χ'' under ac magnetic field of 5 Oe, **g**, dλ'/d$H$ and **h**, dλ''/d$H$ at ac magnetic field of 1 Oe, both in FC process under dc magnetic field of 2 T, with selected frequencies.

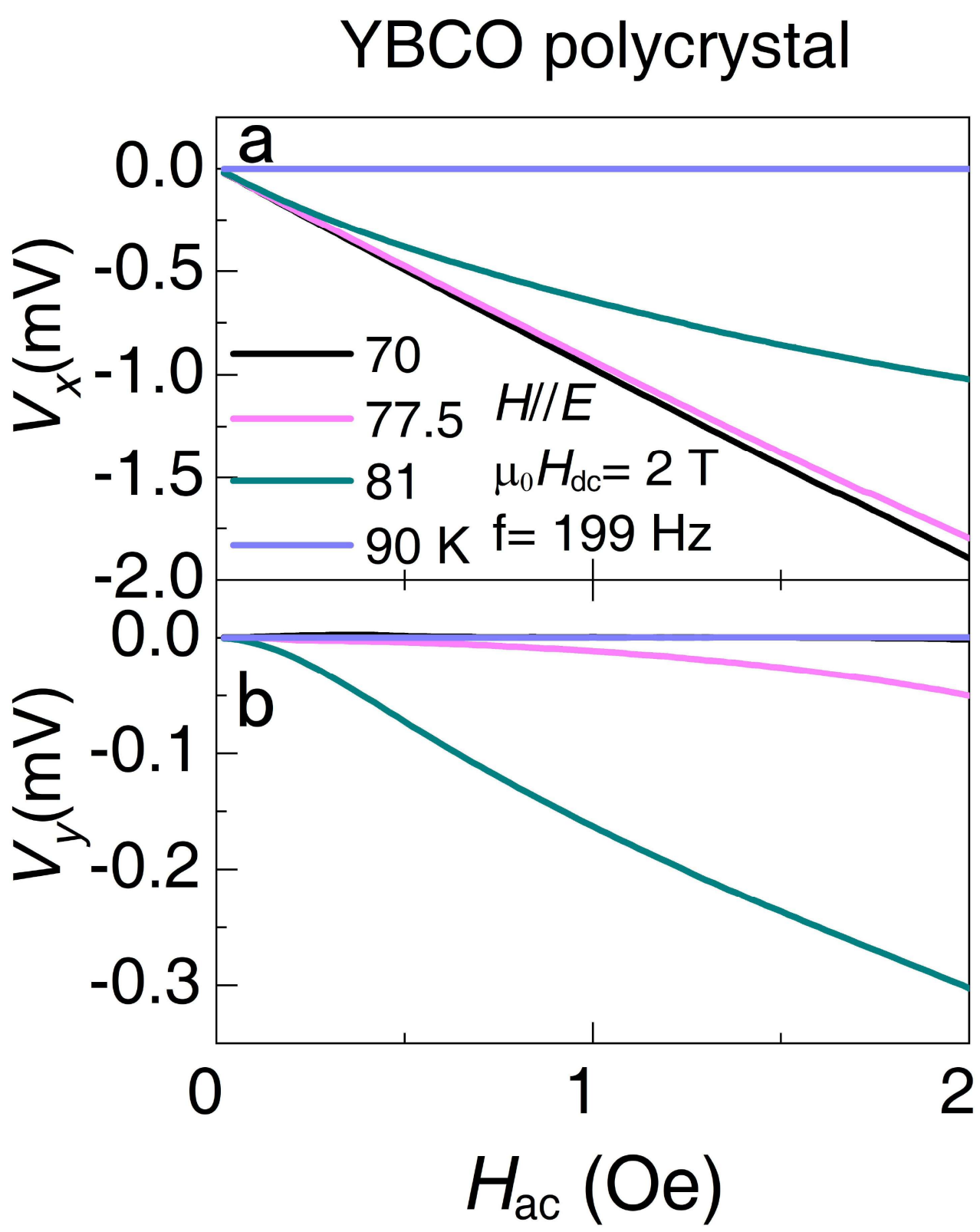

**Supplementary Figure S4.** Under a dc magnetic field of 2 T, the ac ME voltage **a**, $V_x$ **b**, $V_y$ as a function of $H_{ac}$ at selected temperatures of 70, 77.5, 81 and 90 K.

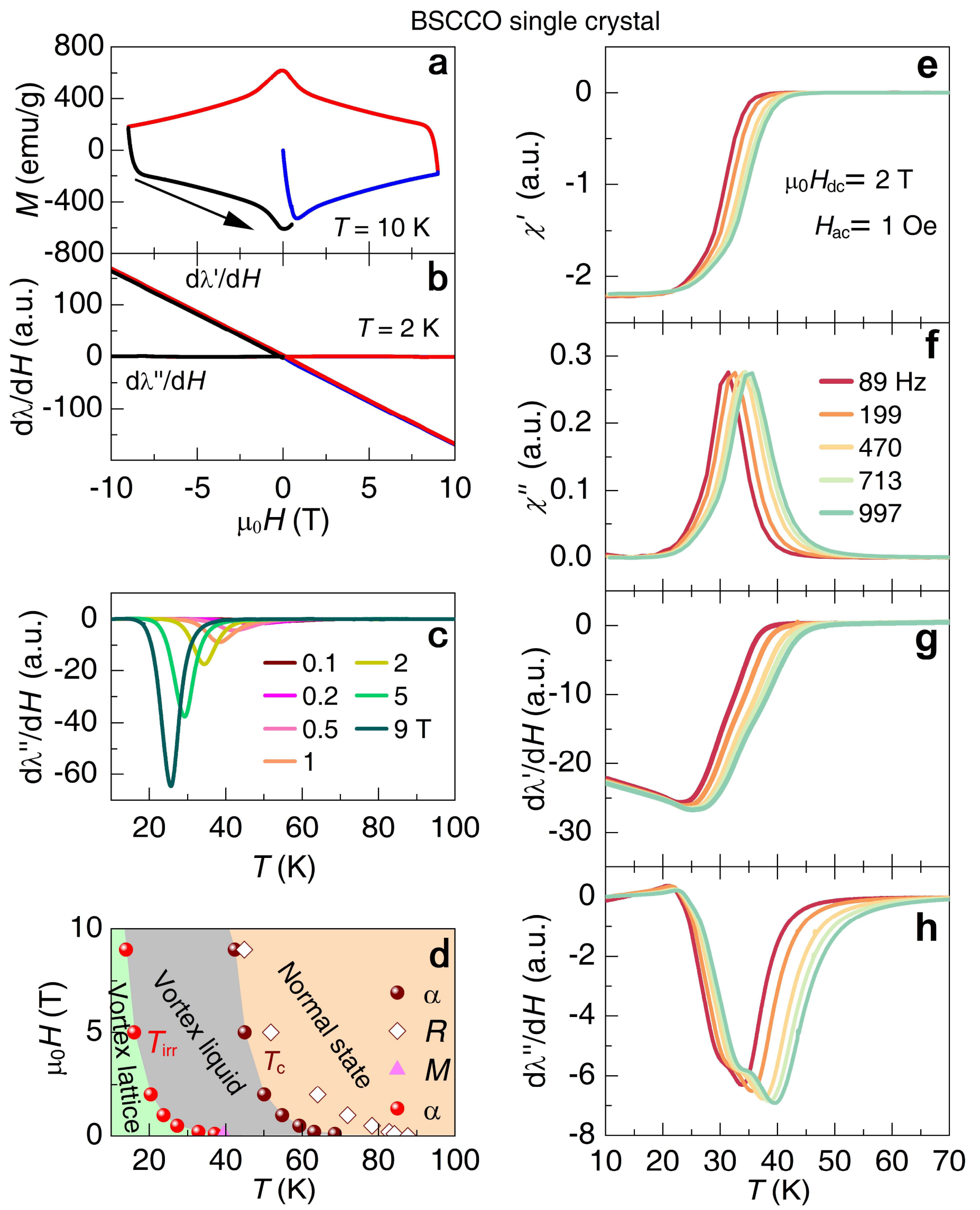

**Supplementary Figure S5. Basic characteristic and *H*-*T* phase diagram of BSCCO single crystal.** Magnetic field dependence of **a**, *M* at 10 K and **b**, dλ'/d*H* and dλ"/d*H* at 2 K. Measurements were performed after ZFC process. **c**, Temperature dependence of dλ"/d*H* in FC process under selected magnetic fields. **d**, The *H*-*T* phase diagram based on the data in panel **c** and Fig. 3**b**, **e** and **h**. Temperature dependent **e**, χ' and **f**, χ", **g**, dλ'/d*H* and **h**, dλ"/d*H* in FC process under dc and ac magnetic fields of 2 T and 1 Oe, with selected frequencies.

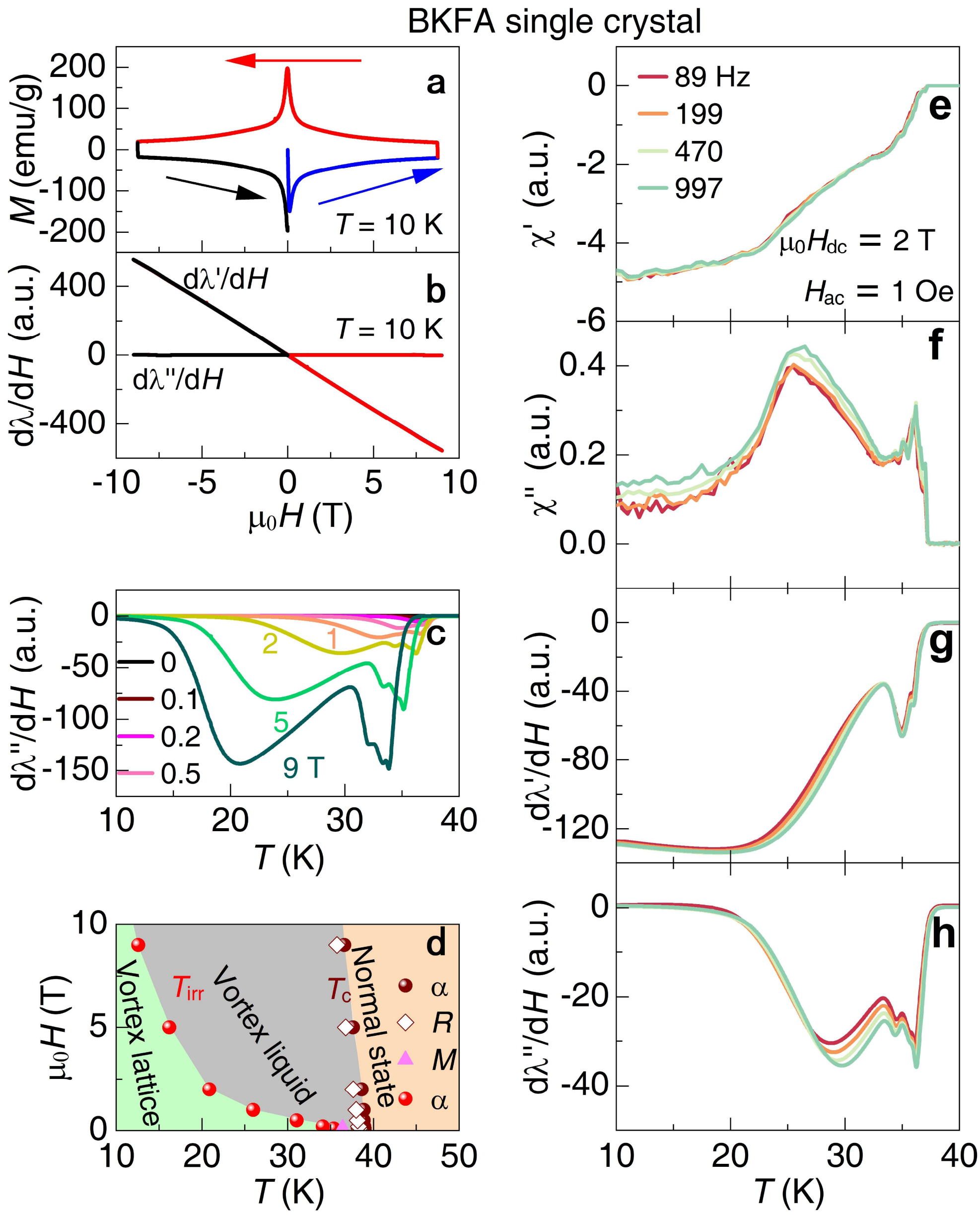

**Supplementary Figure S6. Basic characteristic and *H*-*T* phase diagram of BKFA single crystal.** Magnetic field dependence of **a**, *M* and **b**, dλ'/d*H* and dλ″/d*H* at 10 K. Measurements were performed after ZFC to 10 K. **c**, Temperature dependent dλ''/d*H* in FC process under selected magnetic fields. **d**, The *H*-*T* phase diagram based on the data in panel **c** and Fig. 3**c**, **f** and **i**. Temperature dependent **e**, χ' and **f**, χ'', **g**, dλ'/d*H* and **h**, dλ''/d*H* in FC process under dc and ac magnetic fields of 2 T and 1 Oe, with selected frequencies.

## Specific explanations for $v$, and subsequently $\omega_0$

In a two-dimensional lattice, the sound wave speed is expressed as: $v = \sqrt{Y/\rho}$ (where $Y$ is Young's modulus and ρ is vortex density proportional to $H$) [1]. Meanwhile, it is expected that the lattice vibration ($Y$) of the two-dimensional vortex lattice is mainly governed by shear modulus $C_{66}$, which is proportional to $H$ as well [1]. As a consequence, $v$ and $\omega_0=\pi mv/L$ ($m$ is an integer number and $L$ is sample geometry) will be $H$ independent.

## Specific formula derivation for the dynamic term

In Eq. (2), the dynamic term $\delta\lambda_{ac}/H_{ac}$ can be regarded as a forced oscillation of the geometry $x$ with a weak damping rate $2\beta$:

$$\ddot{x} + 2\beta\dot{x} + \omega_0^2 x = ngH_{ac}sin\omega t \tag{S1}$$

where $\omega_0 >> \omega$, $n$ is the total number of vortices in the vortex lattice, $g$ is the effective acceleration factor generated by each vortex line. Solving this equation gives a steady state solution: $x = \lambda_t \sin(\omega t - \varphi)$ where the amplitude $\delta\lambda_{ac}$ and phase angle φ are:

$$\delta\lambda_{ac} = \sqrt{\frac{g^2n^2H_{ac}^2}{\left(\omega_0^2-\omega^2\right)^2+4\beta^2\omega^2}} \approx \frac{ngH_{ac}}{\omega_0^2} \text{ and } \varphi = \arctan\frac{2\beta\omega}{\omega_0^2-\omega^2} \approx 0 \tag{S2}$$